\documentclass[runningheads]{llncs}
\usepackage[T1]{fontenc}
\usepackage{graphicx}
\usepackage{soul}
\usepackage{url}
\usepackage[hidelinks]{hyperref}
\usepackage[utf8]{inputenc}
\usepackage[small]{caption}
\usepackage{amsmath}
\usepackage{booktabs}
\usepackage{algorithm}
\usepackage{algorithmic}  
\usepackage{todonotes}
\usepackage{comment}

\usepackage{paralist}                                       
\usepackage{xspace}
\usepackage{amssymb}
\usepackage{hyperref}
\usepackage[inline]{enumitem}
\usepackage{breqn}

\newcommand{\ag}{ {\mthfun{ag}} \xspace}
\newcommand{\Env}{ {\mthfun{Env}} \xspace}
\newcommand{\Task}{ {\mthfun{Task}} \xspace}

\newcommand{\Win}{\mathsf{Win}}

\renewcommand\paragraph[1]{{\textit{#1}}}

\newcommand{\A}{\mathcal{A}} 
\newcommand{\C}{\mathcal{C}} \newcommand{\D}{\mathcal{D}}

 \renewcommand{\L}{\mathcal{L}}
\newcommand{\M}{\mathcal{M}} 
 
 \newcommand{\R}{\mathcal{R}}
 
\newcommand{\U}{\mathcal{U}} 
 \newcommand{\X}{\mathcal{X}}
\newcommand{\Y}{\mathcal{Y}}

\newcommand{\limp}{\mathbin{\rightarrow}}

\newcommand{\commentout}[1]{}

\newcommand{\Next}{\raisebox{-0.27ex}{\LARGE$\circ$}}
\newcommand{\Wnext}{\raisebox{-0.27ex}{\LARGE$\bullet$}}
\newcommand{\Until}{\mathop{\U}}

\newcommand{\true}{\mathit{true}}

\newcommand{\false}{\mathit{false}}

\newcommand{\tm}[1]{\ \text{#1}\ }
\newcommand{\tiff}{\tm{iff}}
\newcommand{\trace}{\pi}

\newcommand{\LTL}{{\sc ltl}\xspace}
\newcommand{\LTLf}{{\sc ltl}$_f$\xspace}

\newcommand{\LDLf}{{\sc ldl}$_f$\xspace}

\newcommand{\NFA}{{\sc nfa}\xspace}

\newcommand{\DFA}{{\sc dfa}\xspace}
\newcommand{\DFAs}{{\sc dfa}s\xspace}

\newcommand{\f}{\mathit{f}}

\newcommand{\Nat}{{\rm I\kern-.23em N}}
\newcommand{\Prop}{\P}

\newcommand{\lAND}{\wedge}
\newcommand{\stenv}{\sigma_{\mthsym{env}}}

\newcommand{\stag}{{\sigma}_{\mthsym{\ag}}}

\newcommand{\env}{{\mthfun{env}}}

\newcommand{\DA}{{\sc da}\xspace}

\newcommand{\DAs}{\textsc{da}s\xspace}

\renewcommand{\Prop}{\mthrel{Prop}\xspace}

\newcommand{\reach}{{\operatorname{reach}}\xspace}
\newcommand{\safe}{{\operatorname{safe}}\xspace}

\newcommand{\play}{\mthfun{play}}

\usepackage{xargs}

\usepackage{xstring}

\newcommand{\ignore}[1]{}
\newcommand{\argemp}[2]{\if&#1&\else#2\fi}

\newcommand{\argdef}[2]{\if&#1&#2\else#1\fi}

\newcommand{\argint}[3]{\if&#2&\else#1#2#3\fi}

\newcommand{\argext}[3]{\if&#1&#3\else#1\if&#3&\else#2#3\fi\fi}

\newcommandx{\mthfnt}[3][1=, 2=0]{{
	\IfStrEqCase{#1}
	{%
		{}%
		{#3}%
		{Name}%
		{%
			\IfStrEqCase{#2}
			{%
				{0}{\mathcal{#3}}%
				{1}{\mathscr{#3}}%
				{2}{\mathfrak{#3}}%
				{3}{\mathbb{#3}}%
			}
			[\ensuremath{\clubsuit}]%
		}%
		{Set}%
		{%
			\IfStrEqCase{#2}
			{%
				{0}{\mathrm{#3}}%
				{1}{\mathsf{#3}}%
				{2}{\mathbb{#3}}%
				{3}{\mathbf{#3}}%
			}
			[\ensuremath{\clubsuit}]%
		}%
		{Fun}%
		{%
			\IfStrEqCase{#2}
			{%
				{0}{\mathsf{#3}}%
				{1}{\mathrm{#3}}%
			}
			[\ensuremath{\clubsuit}]%
		}%
		{Rel}%
		{%
			\IfStrEqCase{#2}
			{%
				{0}{\mathit{#3}}%
				{1}{\mathtt{#3}}%
			}
			[\ensuremath{\clubsuit}]%
		}%
		{Sym}%
		{%
			\IfStrEqCase{#2}
			{%
				{0}{\mathtt{#3}}%
				{1}{\mathbf{#3}}%
			}
			[\ensuremath{\clubsuit}]%
		}%
		{Elm}%
		{\mathnormal{#3}}
	}
[\ensuremath{\clubsuit}]%
}}

\newcommand{\mthsub}[1]{\argemp{#1}{\ensuremath{_{\mathnormal{#1}}}}}

\newcommand{\mthsup}[1]{\argemp{#1}{\ensuremath{^{\mathnormal{#1}}}}}

\newcommandx{\mth}[5][1=, 2=0, 4=, 5=]{{\ensuremath{\mthfnt[#1][#2]{#3}\mthsub{#4}\mthsup{#5}}}}

\newcommandx{\mtharg}[6][1=, 2=0, 4=, 5=]{{\mth[#1][#2]{#3}[#4][#5]\ensuremath{\argint{(}{#6}{)}}}}

\newcommand{\mthstyfun}{0}
\newcommand{\mthfun}[1][]{\mth[Fun][\argdef{#1}{\mthstyfun}]}

\newcommand{\mthstyrel}{0}
\newcommand{\mthrel}[1][]{\mth[Rel][\argdef{#1}{\mthstyrel}]}

\newcommand{\mthstysym}{0}
\newcommand{\mthsym}[1][]{\mth[Sym][\argdef{#1}{\mthstysym}]}

\begin{document}

\title{\LTLf Synthesis Under Environment Specifications \\ for Reachability and Safety Properties}

\titlerunning{\LTLf Synthesis Under Env. for Reach. and Safe. Properties}
\author{Benjamin Aminof\inst{1} \and  
  Giuseppe De Giacomo\inst{1,2}\and
  Antonio Di Stasio\inst{2}\and\\
  Hugo Francon\inst{3}\and 
  Sasha Rubin\inst{4}\and
  Shufang Zhu\inst{2} \thanks{All authors are corresponding authors.}}
% %
\authorrunning{Aminof et al.}
\institute{Sapienza University of Rome, Italy\\
\email{benj@forsyte.at}
\and
University of Oxford, UK\\
\email{\{giuseppe.degiacomo, antonio.distasio, shufang.zhu\}@cs.ox.ac.uk}
\and
ENS Rennes, France\\
\email{hugo.francon@ens-rennes.fr}
\and
The University of Sydney, Australia\\
\email{sasha.rubin@sydney.edu.au}}
\maketitle              % typeset the header of the contribution

\newcommand\changed[1]{\textcolor{red}{#1}}

\begin{abstract}
In this paper, we study \LTLf synthesis under environment specifications for arbitrary reachability and safety properties. We consider both kinds of properties for both agent tasks and environment specifications, providing a complete landscape of synthesis algorithms. For each case, we devise a specific algorithm (optimal wrt complexity of the problem) and prove its correctness. The algorithms combine common building blocks in different ways. While some cases are already studied in literature others are studied here for the first time.
\end{abstract}
\section{Introduction}
Synthesis under environment specifications consists of synthesizing an agent strategy (aka plan or program) that realizes a given task against all possible environment responses (i.e., environment strategies). The agent has some indirect knowledge of the possible environment strategies through an environment specification, and it will use such knowledge to its advantage when synthesizing its strategy \cite{DBLP:conf/kr/AminofGMR18,AGMR19,CBM18,PnueliR89}. This problem is tightly related to planning in adversarial nondeterministic domains \cite{GeBo13}, as discussed, e.g., in \cite{CamachoBM19,DeGiacomoS18}.

In this paper, we study synthesis under environment specifications, considering both \emph{agent task specifications} and \emph{environment specifications} expressed in Linear Temporal Logic on finite traces~(\LTLf). These are logics that look at finite traces or finite prefixes of infinite traces. 
For concreteness, we focus on \LTLf \cite{DegVa13,DegVa15}, but the techniques presented here extend immediately to other temporal logics on finite traces, such as Linear Dynamic Logics on finite traces, which is more expressive than \LTLf \cite{DegVa13}, and Pure-Past \LTL, which has the same expressiveness as \LTL but evaluates a trace backward from the current instant~\cite{DDFR20}.  

Linear temporal logics on finite traces provide a nice embodiment of the notable triangle among Logics, Automata, and Games \cite{2001automata}. These logics are full-fledged logics with high expressiveness over finite traces, and they can be translated into classical regular finite state automata; moreover, they can be further converted into deterministic finite state automata (\DFAs). This transformation yields a game represented on a graph. In this game, one can analyze scenarios where the objective is to reach certain final states.
%they can further be by transforming such automata into deterministic finite state automata~(\DFAs) one gets a game on a graph over which one can analyze games with final states as targets. 
Finally, despite the fact that producing a \DFA corresponding to an \LTLf formula can require double-exponential time, the algorithms involved --- generating alternating automata (linear), getting the nondeterministic one (exponential), determinizing it (exponential), solving reachability games (poly) --- are particularly well-behaved from the practical computational point of view~\cite{TaVa20,TabakovV05,ZhuTPV21}.

In this paper, however, we consider \LTLf specifications in two contexts which we denote
as 

% \[\exists\varphi \mbox{ and } \forall\varphi \mbox{ with $\varphi$ an arbitrary \LTLf formula}\]
{\centerline{$\exists\varphi \mbox{ and } \forall\varphi \mbox{ with $\varphi$ an arbitrary \LTLf formula}$.}}

The first one specifies a \emph{reachability} property: there exists a finite prefix $\pi_{<k}$ of an infinite trace $\pi$ such that $\pi_{<k} \models \varphi$. This is the classical use of \LTLf to specify synthesis tasks \cite{DegVa15}. 
The second one specifies a \emph{safety} property:  every finite prefix $\pi_{<k}$ of an infinite trace $\pi$ is such that $\pi_{<k} \models \varphi$. This is the classical use of \LTLf to specify environment behaviours \cite{ADMR18arxiv,DDTVZ21}. 
The formulas $\forall\varphi$ and $\exists\varphi$ with $\varphi$ in \LTLf capture exactly two well-known classes of \LTL properties in Manna and Pnueli's Temporal Hierarchy \cite{MannaPnueli90}. Specifically, $\exists\varphi$ captures the \emph{co-safety properties} and $\forall\varphi$ captures the \emph{safety properties} (in \cite{MannaPnueli90}, expressed respectively as  $\Diamond \psi$ and $\Box\psi$ with $\psi$ an arbitrary Pure-Past \LTL formulas, which consider only past operators.) 

We let $\Env$ and $\Task$ denote an environment specification and a task specification, respectively, consisting of a safety ($\forall\varphi$) and/or reachability property ($\exists\varphi$). This gives rise to 12 possible cases: 3 without any environment specifications, 3 with safety environment specifications ($\forall\varphi$), 3 with reachability environment specifications ($\exists\varphi$), and 3 with both safety and reachability environment specifications ($\exists\varphi\land\forall\varphi$). 
For each of these, we provide an algorithm, which is optimal wrt the complexity of the problem, and prove its correctness. When the problem was already solved in literature, we give appropriate references (e.g., $\Task = \exists\varphi$ and $\Env = true$ is classical \LTLf synthesis, solved in \cite{DegVa15}).
% As a result, we get a complete landscape of algorithms for \LTLf synthesis considering reachability and safety properties for both agent tasks and environment specifications.
In fact, we handle all the cases involving reachability in the environment specifications by providing  a novel algorithm that solves the most general case
of $\Env = \exists \varphi_1 \land \forall \varphi_2$ and $\Task = \exists \varphi_3 \land \forall \varphi_4$.\footnote{In fact, this algorithm can solve all cases, but it’s much more involved compared to the direct algorithms we provide for each case.} 

These algorithms use the common building blocks (combining them in different ways): the construction of the \DFAs of the \LTLf formulas, Cartesian products of such \DFAs, considering these \DFAs as the game arena and solving games for reachability/safety objectives. 
Also, all these problems have a 2EXPTIME-complete complexity. The hardness comes from \LTLf synthesis \cite{DegVa15}, and the membership comes from the \LTLf-to-\DFA construction, which dominates the complexity since computing the Cartesian products and solving reachability/safety games is polynomial.\footnote{For pure-past \LTL,  obtaining the \DFA from a pure-past \LTL formula is single exponential \cite{DDFR20}, and indeed the problems and all our algorithms become EXPTIME-complete.}  
Towards the actual application of our algorithms, we observe that although the \DFAs of \LTLf formulas are worst-case double-exponential, there is empirical evidence showing that the determinization of \NFA, which causes one of the two exponential blow-ups, is often polynomial in the \NFA \cite{TabajaraV19,TabakovV05,ZhuTPV21}. Moreover, in several notable cases, e.g., in all DECLARE patterns \cite{Westergaard11}, the \DFAs are polynomial in the \LTLf formulas, and so are our algorithms.

\section{Preliminaries}\label{sec:pre}

\paragraph{Traces.}
For a finite set  $\Sigma$, let $\Sigma^\omega$ (resp. $\Sigma^+,\Sigma^*$) denote the set of infinite strings (resp. non-empty finite strings, finite strings) over $\Sigma$.  We may write concatenation of sets using $\cdot$, e.g., $\Sigma \cdot \Sigma$ denotes the set of strings over $\Sigma$ of length $2$. The length of a string is denoted $|\pi|$, and may be infinite. Strings are indexed starting at $0$.  For a string $\pi$ and $k \in \Nat$ with $k < |\pi|$, let $\pi_{<k}$ denote the finite prefix of $\pi$ of length $k$.  For example, if $\pi = \pi_0 \pi_1 \ldots \pi_n$, then $|\pi|=n+1$ and $\pi_{<2} = \pi_0 \pi_1$. Typically, $\Sigma$ will be the set of interpretations~(i.e., assignments) over a set $\Prop$ of atomic propositions, i.e., $\Sigma = 2^{\Prop}$. Non-empty strings will also be called \emph{traces}. 

\paragraph{Linear-time temporal logic on finite traces.} % (\LTLftitle)
\LTLf is a variant of Linear-time temporal logic~(\LTL) interpreted over \emph{finite}, instead of infinite, traces \cite{DegVa13}. Given a set $\Prop$
of atomic propositions, \LTLf formulas $\varphi$ are defined by the following grammar:
%%
% \[\begin{array}{rcl}
% \varphi &::=& p \mid \lnot \varphi \mid \varphi \land \varphi \mid \Next\varphi \mid \varphi\Until\varphi
% \end{array}
% \]
%%
$\varphi ::= p \mid \lnot \varphi \mid \varphi \land \varphi \mid \Next\varphi \mid \varphi\Until\varphi$
where $p \in \Prop$ denotes an atomic proposition, $\Next$ is read
\emph{next}, and $\Until$ is read \emph{until}. We abbreviate other Boolean connectives and operators.

For a finite trace $\trace \in (2^\Prop)^+$, an \LTLf formula $\varphi$, and a position $i$ ($0 \leq i < |\trace|$), define $\trace, i \models \varphi$ (read ``$\varphi$ \emph{holds} at position $i$") by induction, as follows:

\begin{compactitem}
    \item 
    $\trace, i \models p \tiff p \in \trace_i\nonumber$ (for $p\in\Prop$);
    \item 
    $\trace, i \models \lnot \varphi \tiff \trace, i \not\models \varphi\nonumber$;
    \item 
    $\trace, i \models \varphi_1 \lAND \varphi_2 \tiff \trace, i \models \varphi_1 \tm{and} \trace, i \models \varphi_2\nonumber$;
    \item 
    $\trace, i \models \Next\varphi \tiff i < |\trace|-1 \tm{and} \trace,i+1 \models \varphi$;
    \item 
    $\trace, i \models \varphi_1 \Until \varphi_2$ iff for some $j~(i\le j < |\trace|)$, $\trace,j \models\varphi_2$, and for all $k~(i\le k < j), \trace, k \models \varphi_1$.
\end{compactitem}

\noindent
We write $\trace \models \varphi$, if $\trace,0 \models \varphi$ and say that $\trace$
\emph{satisfies} $\varphi$. Write $\L(\varphi)$ for the set of finite traces over $\Sigma = 2^{\Prop}$ that satisfy $\varphi$. 
%We use $last$, which stands for $\neg \Next true$. 
In addition, we define the \emph{weak next} operator $\Wnext \varphi \equiv \neg \Next \neg \varphi$.
Note that: $\lnot\Next\varphi$ is not, in general, logically equivalent to $\Next\lnot\varphi$, but we have that $\neg \Next \varphi \equiv \Wnext \neg \varphi$.
%; and $\pi,i \models \Last$ iff $i$ is the final position in the trace~($i=|\trace|-1$).

\paragraph{Domains.}  
A \emph{domain (aka transition system, aka arena)} is a tuple $\D = (\Sigma, Q, \iota, \delta)$, where $\Sigma$ is a finite alphabet, $Q$ is a finite set of states, $\iota \in Q$ is the initial state, $\delta: Q \times \Sigma \to Q$ is a transition function. For an infinite string $w = w_0 w_1 w_2 \ldots \in \Sigma^\omega$ a \emph{run} of $\D$ on $w$ is a sequence $r = q_0 q_1 q_2 \ldots \in Q^\omega$ that $q_0 = \iota$ and $q_{i+1} \in \delta(q_i, w_i)$ for every $i$ with $0 \leq i$. A \emph{run} of $\D$ on a finite string $w = w_0 w_1 \ldots w_n$ over $\Sigma$ is a sequence $q_0 q_1 \cdots q_{n+1}$ such that $q_0 = \iota$ and $q_{i+1} \in \delta(q_i,w_i)$ for every $i$ with $0 \leq i < n+1$. Note that every string has exactly one run of $\D$. 

\paragraph{Deterministic finite automaton (DFA).}
A \emph{DFA} is a tuple $\M = (\D,F)$ where $\D$ is a domain and $F \subseteq Q$ is a set of \emph{final} states.
A finite word $w$ over $\Sigma$ is \emph{accepted} by $\M$ if the run of $\M$ on $w$ ends in a state of $F$. 
The set of all such finite strings is denoted $\L(\M)$, and is called the \emph{language} of $\M$.

\begin{theorem}
\label{ltlf2dfa}\cite{DegVa15}  
Every \LTLf formula $\varphi$ over atoms $\Prop$ can be translated into a \DFA $\M_\varphi$ over alphabet $\Sigma = 2^{\Prop}$ such that for every finite string $\pi$ we have that $\pi \in \L(\M)$ iff $\pi \models \varphi$. This translation takes time double-exponential in the size of $\varphi$.
\end{theorem}

\paragraph{Properties of infinite strings.}
A \emph{property} is a set $P$ of infinite strings over $\Sigma$, i.e., $P \subseteq \Sigma^{\omega}$. 
We say that $P$ is a \emph{reachability} property if there exists a set $T \subseteq \Sigma^+$ of finite traces such that if $w  \in P$ then some finite prefix of $w$ is in $T$. We say that $P$ is a \emph{safety} property if there exists a set $T \subseteq \Sigma^+$ of finite traces such that if $w \in P$, then every finite prefix of $w$ is in $T$. It is worth noting that the complement of a reachability property is a safety property, and vice versa.

An \LTLf formula can be used to denote a reachability (resp.,  safety) property over $\Sigma = 2^\Prop$ as follows. 
\begin{definition}
For an \LTLf formula $\varphi$, let $\exists \varphi$ denote set of traces $\trace$ such that some finite prefix of $\pi$ satisfies $\varphi$, and let $\forall \varphi$ denote set of traces $\trace$ such that every finite (non-empty) prefix of $\trace$ satisfies $\varphi$.
\end{definition}
Note that $\exists \varphi$ denotes a reachability property, and $\forall \varphi$ denotes a safety property. From now on, ``prefix" will mean ``finite non-empty prefix".
Note also that for an \LTLf formula, $\L(\varphi)$ is a set of finite traces. On the other hand, $\L(\exists \varphi)$ (and similarly $\L(\psi)$ where $\psi$ is a Boolean combination of formulas of the form $\exists \varphi$ for \LTLf formulas $\varphi$) is a set of infinite traces.  
\begin{comment}
For example, let $\varphi_1,\varphi_2$ be \LTLf formulas. 

\begin{compactitem}
    \item 
    $\pi \models \exists \varphi_1 \land \forall \varphi_2$ iff some prefix of $\pi$ satisfies $\varphi_1$ and all prefixes of $\pi$ satisfy $\varphi_2$. 
    \item 
    $\pi \models \exists (\varphi_1 \land \varphi_2)$ iff some prefix of $\pi$ satisfies both $\varphi_1$ and $\varphi_2$.
\end{compactitem}
\end{comment}
In this paper, we consider $\exists \varphi$, $\forall \varphi$, and $\exists \varphi \land \forall \varphi$ to specify both agent tasks and environment behaviours. 

\begin{comment}
\begin{remark} 
Linear temporal logic (\LTL)~\cite{Pnueli77} has the same syntax as \LTLf, and its semantics is over infinite traces. Thus, \LTL formulas denote properties. In particular, every reachability (resp. safety) property expressible in \LTL is expressible as $\exists \varphi$ (resp. $\forall \varphi$) for some \LTLf formula $\varphi$~\cite{DDTVZ21}.
\end{remark}
\end{comment}

\paragraph{Deterministic automata on infinite strings (\DA).}
%Automata over infinite strings.}
Following the automata-theoretic approach in formal methods, we will compile formulas to automata. We have already seen that we can compile \LTLf formulas to \DFAs. We now introduce automata over infinite words to handle certain properties of infinite words.
A \emph{deterministic automaton} (\DA, for short) is a tuple $\A = (\D,\alpha)$ where $\D$ is a transition system, say with the state set $Q$, and $\alpha \subseteq Q^\omega$ is called an \emph{acceptance condition}. An infinite string $w$ is \emph{accepted} by $\A$ if its run is in $\alpha$. The set of all such infinite strings is denoted $\L(\A)$, and is called the \emph{language} of~$\A$.

We consider reachability~($\reach$) and safety~($\safe$) acceptance conditions, parameterized by a set of target states $T \subseteq Q$:
\begin{compactitem}
\item 
    $\reach(T) = \{ q_0 q_1 q_2 \ldots \in Q^\omega  \mid \exists k \geq 0: q_k \in T\}$. In this case, we call $\A$ a \emph{reachability automaton}. 
    \item 
    $\safe( T)= \{ q_0 q_1 q_2 \ldots \in Q^\omega \mid \forall k \geq 0: q_k \in T\}$. In this case, we call $\A$ a \emph{safety automaton}.

\end{compactitem} 

\begin{remark}
Every reachability~(resp.~safety) property expressible in \LTL is the language of a reachability automaton (resp. safety automaton)~\cite{DegVa13,KV01,RS59}.
\end{remark}

\section{Problem Description}\label{sec:problem_def}

\paragraph{Reactive Synthesis.}
Reactive Synthesis (aka Church's Synthesis) is the problem of turning a specification of an agent's task and of its environment into a strategy (aka policy). This strategy can be employed by the agent to achieve its task, regardless of how the environment behaves. In this framework, the agent and the environment are considered players in a turn-based game, in which players move by picking an evaluation of the propositions they control.
Thus, we partition the set $\Prop$ of propositions into two disjoint sets of propositions $\X$ and $\Y$, and with a little abuse of notation, we denote such a partition as $\Prop = \Y \cup \X$. Intuitively, the propositions in $\X$ are controlled by the environment, and those in $\Y$ are controlled by the agent. In this work (in contrast to the usual setting of reactive synthesis), the agent moves first. The agent moves by selecting an element of $2^\Y$, and the environment responds by selecting an element of $2^\X$. This is repeated forever, and results in an infinite trace (aka play). From now on, unless specified otherwise, we let $\Sigma = 2^{\Prop}$ and $\Prop = \Y \cup \X$.  We remark that the games considered in this paper are games of perfect information with deterministic strategies.

An \emph{agent strategy} is a function $\sigma_\ag:(2^{\X})^* \to 2^{\Y}$. An environment strategy is a function $\sigma_\env:(2^\Y)^+ \to 2^\X$. A strategy $\sigma$ is \emph{finite-state} (aka \emph{finite-memory}) if it can be represented as a finite-state input/output automaton that, on reading an element $h$ of the domain of $\sigma$, outputs the action $\sigma(h)$.  
A trace $\pi = (Y_0 \cup X_0) (Y_1 \cup X_1) \dots \in (2^{\Y \cup \X})^\omega$ \emph{follows an agent strategy} $\sigma_\ag: (2^\X)^* \rightarrow 2^\Y$ if $Y_0 = \sigma_\ag(\epsilon)$ and $Y_{i+1} = \sigma_\ag(X_0X_1\ldots X_i)$ for every $i \geq 0$, and it \emph{follows an environment strategy} $\sigma_\env$ if $X_i = \sigma_\env(Y_0 Y_1 \ldots Y_i)$ for all $i \geq 0$. We denote the unique infinite sequence~(play) that follows $\sigma_\ag$ and $\sigma_\env$ as $\play( \stag,\stenv)$. 
Let $P$ be a property over the alphabet $\Sigma = 2^\Prop$, specified by formula or \DA. An agent strategy $\sigma_\ag$~(resp., environment strategy $\sigma_\env$) \emph{enforces} $P$ if for every environment strategy $\sigma_\env$ (resp., agent strategy $\sigma_\ag$), we have that $\play(\sigma_\ag,\sigma_\env)$ is in $P$. In this case, we write $\sigma_{\ag} \rhd P$ (resp. $\sigma_\env \rhd P$). We say that $P$ is \emph{agent (resp., environment) realizable} if there is an agent (resp. environment) strategy that enforces $P$. %We consider two computational problems. 

\begin{comment}
\begin{definition}[Synthesis without Environment specifications] \label{dfn:synthesis}
The \emph{realizability problem} asks whether a given property $P$ is agent realizable. The \emph{synthesis problem} asks, given a property $P$, to decide if there is an agent strategy that enforces $P$, and if so, to return a finite-state strategy that does.
\end{definition}
\end{comment}

\paragraph{Synthesis under Environment Specifications.} Typically, an agent has some knowledge of how the environment works, represented as a fully observable model of the environment, which it can exploit to enforce its task~\cite{AGMR19}. 
Formally, let $\Env$ and $\Task$ be properties over alphabet $\Sigma = 2^{\Prop}$, denoting the environment specification and the agent task, respectively. 

Note that while the agent task $\Task$ denotes the set of desirable traces from the agent's perspective, the environment specification $\Env$ denotes the set of environment strategies that describe how the environment reacts to the agent’s actions (no matter what the agent does) in order to enforce $\Env$. Specifically, $\Env$ is treated as a set of traces when we reduce the problem of synthesis under environment specification to standard reactive synthesis.

We require a consistency condition of $\Env$, i.e., there must exist at least one environment strategy $\sigma_\env \rhd \Env$. 
An agent strategy $\sigma_\ag$ enforces $\Task$ \emph{under the environment specification} $\Env$, written 
$\sigma_\ag \ \rhd_\Env \Task$, if for all $\stenv \ \rhd \Env$ we have that $\play(\sigma_\ag,\stenv)\models \Task$. Note that if $\Env = \true$ then this just says that $\sigma_\ag$ enforces $\Task$ (i.e., the environment specification is missing).

%Here are the computational problems for synthesis under environment specifications.

\begin{definition}[Synthesis under environment specifications]
Let $\Env$ and $\Task$ be properties over alphabet $\Sigma = 2^{\Prop}$, 
% where $\Prop = \X \cup \Y$, 
denoting the environment specification and the agent task, respectively. 
\begin{enumerate*}[label=(\roman*)]
\item The \emph{realizability under environment specifications problem} asks, given $\Task$ and $\Env$, to decide if there exists an agent strategy enforcing $\Task$ under the environment specification $\Env$.

\item The \emph{synthesis under environment specifications problem} asks, given $\Task$ and $\Env$, to return a finite-state agent strategy enforcing $\Task$ under the environment specification $\Env$, or say that none exists.
\end{enumerate*} 
\end{definition}

In \cite{AGMR19} is shown that for any linear-time property\footnote{Technically, the properties should be Borel, which all our properties are.}, synthesis under environment specifications can be reduced to synthesis without environment specifications. 
\begin{comment}
\begin{theorem}[\cite{AGMR19}]\label{thm:synthesis-implication}
Let $\Env, \Task$ be an environment specification and an agent task, respectively. For every agent strategy $\sigma_\ag$, we have that
$\sigma_\ag \rhd (\Env \rightarrow \Task)$ implies $\sigma_\ag \ \rhd_\Env \Task$. Although the converse may fail (there are $\Env,\Task,\sigma_\ag$ such that $\sigma_\ag \ \rhd_\Env \Task$ but $\sigma_\ag \not \rhd (\Env \rightarrow \Task)$), we do have that the following are equivalent:
\begin{enumerate}
    \item 
    There exists $\sigma_\ag$ such that $\sigma_\ag \ \rhd_\Env \Task$.
    \item 
    There exists $\sigma_\ag$ such that $\sigma_\ag \rhd (\Env \rightarrow \Task)$.
\end{enumerate}
\end{theorem}
\end{comment}
%
Thus, in order to show that $\Task$ is realizable under $\Env$ it is sufficient to show that $\Env \limp \Task$ is realizable. Moreover, to solve the synthesis problem for $\Task$ under $\Env$, it is enough to return a strategy that enforces $\Env \limp \Task$.

\begin{table}[!ht]
\centering
\begin{tabular}{|c|c|c|}
\hline
\Task & \Env & Alg. \\
\hline
$\exists \varphi$ & $\true$ & Alg. \ref{alg:reach}\\
$\forall \varphi$ & $\true$ & Alg. \ref{alg:safe}\\
$\exists \varphi_1 \land \forall \varphi_2$ & $\true$  & Alg. \ref{alg:reachsafe}\\
\hline
$\exists \varphi_1$ & $\forall \varphi_2$  & Alg. \ref{alg:safe->reach}\\
$\forall \varphi_1$ & $\forall \varphi_2$  & Alg. \ref{alg:safe->safe}\\
$\exists \varphi_1 \wedge \forall \varphi_2$ & $\forall \varphi_3$  & Alg. \ref{alg:safe->reach+safe}\\
\hline
$\exists \varphi_1 \wedge \forall \varphi_2$ & $\exists \varphi_3 \wedge \forall \varphi_4$  & Alg. 7 \\
\hline
\end{tabular}
\vspace{0.5cm}
\caption{\Task and \Env considered. Note that, from Alg. 7 we get the remaining cases involving reachability environment specifications by suitably setting $ \varphi_1, \varphi_2,\varphi_4$ to $true$.}
\label{algs}
\end{table}
\vspace{-0.6cm}

In the rest of the paper, we provide a landscape of algorithms for \LTLf synthesis considering reachability and safety properties for both agent tasks and environment specifications. However, these synthesis problems are complex and challenging due to the combination of reachability and safety properties.
To tackle this issue, one possible approach is to reduce \LTLf synthesis problems to \LTL synthesis problems through suitable translations, e.g., ~\cite{DDPZ21,DDVZ20,ZTLPV17,ZGPV20}. However, there is currently no methodology for performing such translations when considering combinations of reachability and safety properties.\footnote{In~\cite{CBM18} is shown that the case of \LTLf synthesis under safety and reachability properties can be solved by reducing to games on infinite-word automata. This certain case is covered in our paper, nevertheless, we provide a direct approach that only involves games on finite-word automata.}Additionally, synthesis algorithms for \LTL specifications are generally more challenging than those for \LTLf specifications, both theoretically and practically~\cite{DDTVZ21,DDVZ20,ZGPV20,ZTLPV17}. In this paper, we show that for certain combinations, we can avoid the detour to \LTL synthesis and keep the simplicity of \LTLf synthesis. Specifically, we consider that $\Task$ and $\Env$ take the following forms: $\exists \varphi_1, \forall \varphi_1, \exists \varphi_1 \land \forall \varphi_2$ where the $\varphi_i$ are \LTLf formulas, and in addition we consider the case of no environment specification (formally, $\Env = \true$). This results in 12 combinations.  Algorithms 1-7, listed in Table \ref{algs}, optimally solve all the combinations. All these algorithms adopt some common building blocks while linking them in different ways. 
\begin{theorem}
Let each of \Task and \Env be of the forms $\forall \varphi$, $\exists \varphi$, or $\exists \varphi_1 \land \forall \varphi_2$. Solving synthesis for an agent \Task under  environment specification \Env  is 2EXPTIME-complete. 
\end{theorem}

\section{Building Blocks for the Algorithms}

In this section, we describe the building blocks we will use to devise the algorithms for the problem described in the previous section.

\paragraph{DAs for $\exists \varphi$ and $\forall \varphi$.}
Here, we show how to build the \DA whose language is exactly the infinite traces satisfying $\exists \varphi$ (resp. $\forall \varphi$).  The first step is to convert the \LTLf formula $\varphi$ into a \DFA $\M_\varphi = (\Sigma, Q, \iota, \delta, F)$ that accepts exactly the \emph{finite} traces that satisfy $\varphi$ as in Theorem~\ref{ltlf2dfa}. Then, to obtain a \DA $\A_{\exists \varphi}$ for $\exists \varphi$ define $\A_{\exists \varphi} = (2^{\X \cup \Y}, Q, \iota, \delta, \reach(F))$. It is immediate that $\L(\exists \varphi) = \L(\A_{\exists \varphi})$.
To obtain a  \DA  $\A_{\forall \varphi}$  for $\forall \varphi$  define  $\A_{\forall \varphi} =(2^{\X \cup \Y}, Q , \iota, \delta, \safe(F \cup \{\iota\}))$.
%where $\delta'$ is defined as follows:
%\[
%\delta'(q, Z) = 
%		\begin{cases}
%            \delta(q,Z) &\text{ if $\delta(q,Z) \in T$} \\
%			\bot &\text{ if $q = \bot$ or $\delta(q,Z) \notin T$ }. 
%	       \end{cases}
%\]

The reason $\iota$ is considered a part of the safe set is that the \DFA $\M_{\varphi}$ does not accept the empty string since the semantics of \LTLf precludes this. It is immediate that $\L(\forall \varphi) = \L(\A_{\forall \varphi})$. For $\psi \in \{\exists \varphi, \forall \varphi\}$, we let $\textsc{ConvertDA}(\psi)$ denote the resulting \DA.

%We summarise this as follows:
\begin{lemma}\label{lem:autconstruction}
Let $\varphi$ be an \LTLf formula, and let $\psi \in \{\exists \varphi, \forall \varphi\}$. Then the languages
$\L(\psi)$ and $\L(\textsc{ConvertDA}(\psi))$ are equal.
\end{lemma}

{For formulas of the form $\forall \varphi$ we will suppress the initial state in the objective and so $\textsc{ConvertDA}(\forall \varphi)$ will be written $(\D_{\forall \varphi},\safe(T))$, i.e., $T$ contains $\iota$.}

\paragraph{Games over \DA.} 
The synthesis problems we consider in this paper are solved by reducing them to two-player games. 
We will represent games by \DAs $\A = (\D,\alpha)$ where $\D$ is a transition system, sometimes called an `arena', and $\alpha$ is an acceptance condition, sometimes called a `winning condition'. The game is played between an \emph{agent} (controlling $\Y$) and \emph{environment} (controlling $\X$). 
%Intuitively, the reduction guarantees that the given synthesis problem is realizable (i.e., there exists an agent strategy $\sigma$ that enforces $\Task$ under environment specification $\Env$) iff the agent has a strategy in the two-player game to enforce $\L(\A)$, and we can indeed compute $\sigma_\ag$ from the agent strategy computed from the game.
Intuitively, a position in the game is a state $q \in Q$. The initial position is $\iota$. 
From each position, first the agent moves by setting $Y \in 2^\Y$, then the environment moves by setting $X \in 2^\X$, and the next position is updated to the state $\delta(q,Y \cup X)$. 
%The agent makes a move by looking at the history $h \in Q \cdot (2^\Y \cdot 2^\X \cdot Q)^*$ leading to the current state, and the environment by looking at both the history $h$ and the last agent move $Y \in 2^\Y$. 
This interaction results in an infinite run in $\D$, and the agent is declared the winner if the run is in $\alpha$ (otherwise, the environment is declared the winner).

\begin{definition} \label{dfn:agent strategy wins game}
An agent strategy $\sigma_\ag$ is said to \emph{win the game $(\D,\alpha)$} if for every trace $\pi$ that follows $\sigma_\ag$, the run in $\D$ of $\pi$ is in $\alpha$.
\end{definition}
In other words, $\sigma_\ag$ wins the game if every trace $\pi$ that follows $\sigma_\ag$ is in $L(\D,\alpha)$.
For $q \in Q$, let $\D_q$ denote the transition system $\D$ with initial state $q$, i.e., $\D_q = (\Sigma, Q, q, \delta)$. We say that $q$ is a \emph{winning state} for the agent if there is an agent strategy that wins the game $(\D_q,\alpha)$; in this case, the strategy is said to \emph{win starting from $q$}.

In the simplest settings, we represent agent strategies as functions of the form $f_\ag: Q \to 2^\Y$, called positional strategies. An agent positional strategy $f_\ag$ induces an agent strategy, $\sigma_\ag = \textsc{Strategy}(\D_q,f_{\ag})$, as follows: define $\sigma_\ag(\epsilon) = f_\ag(q)$, and for every finite trace $\pi$ let $\rho$ be the run of $\D_q$ on $\pi$ (i.e., starting in state $q$), and define $\sigma_\ag(\pi) = f_\ag(q')$ where $q'$ is the last state in $\rho$ (i.e., $q' = \rho_{|\pi|}$). In more complex settings, e.g., in the Algorithm \ref{alg:reach+safe->reach+safe}, we will construct functions of the form $f_\ag: Q \cdot (2^\Y \cdot 2^\X \cdot Q)^* \to 2^\Y$, which similarly induce agent strategies $\textsc{Strategy}(\D_q,f_\ag )$ where for every finite trace $\pi = Y_0 \cup X_0, \cdots, Y_k \cup X_k$, and run $q_0, \cdots, q_{k+1}$ of $\pi$ in $\D_q$, define $\sigma_\ag(\pi) = f_\ag(q_0, Y_0 \cup X_0, q_1, Y_1 \cup X_1, \cdots, q_{k+1})$.  {Below the agent strategy $\sigma_\ag = \textsc{Strategy}(\D_q,f_{\ag})$ returned by the various algorithms will be finite state, in the sense that it is representable as a transducer. % obtained $(\D_q,\f_\ag)$. 
For simplicity, with a little abuse of notation, we will return directly $\sigma_\ag$, instead of its finite representation as a transducer.}

Dual definitions can be given for the environment, with the only notable difference being that $f_\env:Q \times 2^X \to 2^\Y$ since the moves of the environment depend also on the last move of the agent (since the agent moves first).

In this paper, besides the terms `environment' and `agent', we also consider the terms `protagonist' and `antagonist'. 
If the \DA $(\D,\alpha)$ is a specification for the agent, then the agent is called the protagonist and the environment is called the antagonist. On the other hand, if the \DA $(\D,\alpha)$ is a specification for the environment, then the environment is called the protagonist, and the agent is called the antagonist.
Intuitively, the protagonist is trying to make sure that the generated traces are in $\L(\D, \alpha)$, and the antagonist to make sure that the generated traces are not in $\L(\D, \alpha)$. Define $\Win_p$ (resp. $\Win_a$) as the set of states $q \in Q$ such that $q$ is a protagonist (resp. antagonist) winning state. 
This set is called protagonist's (resp. antagonist) \emph{winning region}. In this paper, all our games (including reachability and safety games) are \emph{determined}. Therefore:
\begin{lemma} \label{lemma:determined}
For every state $q \in Q$, it holds that $q \in \Win_p$ iff $q \notin \Win_a$. 
\end{lemma}
The problem of \emph{solving} a game $(\D, \alpha)$ for the protagonist is to compute the winning region $\Win_p$ and a function $f_p$ such that $\textsc{Strategy}(\D,f_p)$ wins from every state in $\Win_p$.\footnote{Since strategies can depend on the history, and thus on the starting state in particular, there is always a strategy that wins from every state in $\Win_p$.} To do this, we will also sometimes compute a winning strategy for the antagonist (that wins starting in its winning region).

\paragraph{Solving reachability games and safety games.} We repeatedly make use of solutions to reachability games and safety games given by \DAs $\A$. {Thus, for a protagonist $p \in \{\ag,\env\}$ let $\textsc{Solve}_p(\A)$ denote the procedure for solving the game $\A$, i.e., $p$ is trying to ensure the play is in $\L(\A)$; this procedure returns the protagonist's winning region $\Win_p$ and a function $f_p$ such that $\textsc{Strategy}(\D,f_p)$ wins starting from every state in $\Win_p$~\cite{fijalkow2023games}.}

\paragraph{Product of Transition Systems.} Let $\D_i$ ($1 \leq i \leq k$)  be transition systems over alphabet $\Sigma$. Their \emph{product}, denoted 
$
\textsc{Product}(\D_1,\cdots,\D_k),
$
is the transition system $\D = (\Sigma,Q,\iota,\delta)$ defined as follows:
\begin{enumerate*}[label=(\roman*)]
\item The alphabet is $\Sigma$.
\item The state set is $Q = Q_1 \times \cdots \times Q_k$.
\item The initial state is $\iota = (\iota_1, \cdots, \iota_k)$.
\item The transition function $\delta$ maps a state $(q_1,\cdots,q_k)$ on input $z \in \Sigma$ to the state $(q'_1,\cdots,q'_k)$ where $q'_i = \delta_i(q_i,z)$ ($1 \leq i \leq k$).
\end{enumerate*}
Also, the \emph{lift} of a set $F_i \subseteq Q_i$ to $\D$ is the set $\{(q_1,\cdots,q_k) : q_i \in F_i\} \subseteq Q$.

% \begin{lemma} \label{lem:lift strategies to product}
%   A trace in $\D = \textsc{Product}(\D_1,\cdots,\D_k)$ follows the lift of a strategy $f_\ag/f_\env$ on $\D_i$ to $\D$ if and only if the trace follows the strategy  $f_\ag/f_\env$ on $\D$, for some $i \in \{1,\ldots,k\}$.
% \end{lemma} 

\paragraph{Restriction of a transition system.} The restriction of a transition system, defined as the procedure $\textsc{Restriction}(\D, S)$, restricts $\D = (\Sigma, Q, \iota, \delta)$ to $S \subseteq Q$ is the transition system $\D'= (\Sigma, S \cup \{sink\}, \iota, \delta', \alpha')$ where for all $z \in \Sigma$, $\delta'(sink,z) = sink$, $\delta'(q,z) = \delta(q,z)$ if $\delta(q,z) \in S$, and $\delta'(q,z) = sink$ otherwise. Intuitively, $\D'$ redirect all transitions from $S$ that leave $S$ to a fresh $sink$ state. We may denote the sink by $\bot$.\footnote{
We remark that (i) when we restrict the transition system of a \DA $(\D,\alpha)$ we may need to revise the winning-condition $\alpha$ to express whether reaching $sink$ is good for the protagonist or not (although many times it is not, e.g., when restricting to the winning-region for a safety condition); (ii) in one case, in Algorithm \ref{alg:reach+safe->reach+safe}, we will add two sink states.
}
 
\begin{comment}
\begin{remark}\label{rm:allsafestrategies} 
Given a transition system $\D$ with state set $Q$, and a set $T \subseteq Q$, consider the safety \DA $\A = (\D,\safe,T)$. Let $\Win_p$ be the winning region of the protagonist. Consider the restricted transition system $\D' := \textsc{Restrict}(\D, \Win_p)$. Note that $\A' = (\D',\safe,T)$ is a well-defined \DA (i.e., reaching the $sink$ violates the safety condition). A strategy for the protagonist is winning in $\A$ iff it is winning in $\A'$. Intuitively, this is because winning strategies for a safety condition cannot leave the winning region, and this is the only requirement for them to be winning. Thus, intuitively, the restriction $\D'$ represents all the strategies that enforce $\L(\D,\safe,T)$~\cite{BernetJW02}.
\end{remark}
\end{comment}

\section{Reachability Tasks, No Env Spec}

Algorithm \ref{alg:reach} solves the realizability and synthesis for the case of reachability tasks and no environment specification. Formally, $\Task$ is of the form $\exists \varphi$ where $\varphi$ is an \LTLf formula, and $\Env = \true$. 
This problem is solved in \cite{DegVa15}, but here we rephrase the problem in our notation.
%\vspace{-0.4cm}
\begin{algorithm}[h]
\caption{$\Task = \exists \varphi, \Env = \true$}\label{alg:reach}
\begin{algorithmic}[1]
\REQUIRE \LTLf formula $\varphi$
\ENSURE agent strategy $\sigma_\ag$ that enforces $\exists \varphi$
\STATE $\A = \textsc{ConvertDA}(\exists \varphi)$, say $\A = (\D_{\exists \varphi}, \reach(T))$
\STATE $(W, f_\ag)= $  \textsc{Solve$_{ag}$}$(\A)$
\STATE \textbf{if} {$\iota \not\in W$} \textbf{return} ``Unrealisable" \textbf{endif}
%\IF {$\iota \not\in W$} \RETURN ``Unrealisable" \IENDIF
\STATE \textbf{return} $\sigma_\ag =$ \textsc{Strategy}$(\D_{\exists \varphi},f_\ag)$ 
%\RETURN \sigma_\ag 
\end{algorithmic}
\end{algorithm}
\vspace{-0.4cm}

%\begin{theorem}\label{thm:alg-reach}

%Let $\Task = \exists \varphi,\Env = \true$, where $\varphi$ is an \LTLf formula, and $f_\ag$ the game strategy %returned by Algorithm~\ref{alg:reach}. Then $\sigma_\ag=\textsc{Strategy}(f_\ag,\D_{\exists \varphi})$ is a %strategy that solves synthesis under environment specifications problem.
%\end{theorem}

\begin{theorem}\label{thm:alg-reach}

Algorithm~\ref{alg:reach} solves the synthesis under environment specifications problem with $\Task = \exists \varphi,\Env = \true$, where $\varphi$ is an \LTLf formula.
\end{theorem}
\begin{comment}
\begin{proof}
By Lemma~\ref{lem:autconstruction}, the language of the \DA in Line 1 is $\L(\exists \varphi)$. 
By Lemma~\ref{lem:games-synthesis} it is enough to find check $\iota$ is a winning state, and in this case to compute a winning strategy $f_\ag$ and return $\textsc{Strategy}(f_\ag)$. But finding the winning region and a winning strategy (for every state in the winning region) for reachability games is done in Line 2 using Lemma~\ref{lem:solvereach}.
\end{proof}
\end{comment}
%Note that, with a little abuse of notation we are returnin direclty $\sigma_\ag$

\section{Safety Tasks, No Env Spec} \label{Task=safe, env=none}

Algorithm \ref{alg:safe} handles the case $\Task$ is of the form $\forall \varphi$ where $\varphi$ is an \LTLf formula, and $\Env = \true$. We can use the result in \cite{DegVa15} to solve the synthesis for $\forall \varphi$  from the point of view of the environment.

%\vspace{-0.4cm}
\begin{algorithm}[H]
\caption{$\Task = \forall \varphi, \Env = true$}
\label{alg:safe}
\textbf{Input}: \LTLf formula $\varphi$\\
\textbf{Output}: agent strategy $\sigma_\ag$ that enforces $\forall \varphi$

\begin{algorithmic}[1]
\STATE $\A_1 =  \textsc{ConvertDA}(\forall \varphi)$, say $\A_1 = (\D_{\forall \varphi},\safe(T_1))$
\STATE $(S_1,f_\ag)=  \textsc{Solve}_{ag}(\A_1)$ %\D_{\forall \varphi}, \safe(T_1))$
\STATE \textbf{if} {$\iota \not\in S_1$} \textbf{return} ``Unrealisable" \textbf{endif}

%\IF {$\iota \not\in S_1$}
%\RETURN ``Unrealisable"
%\ENDIF
%\STATE $\sigma_\ag = \textsc{Strategy}(f_\ag)$
\RETURN  $\sigma_\ag = \textsc{Strategy}(\D_{\forall \varphi}, f_\ag)$
\end{algorithmic}
\end{algorithm}
\vspace{-0.4cm}

\begin{theorem}\label{thm:alg-safety}
Algorithm~\ref{alg:safe} solves the synthesis under environment specifications problem with $\Task = \forall \varphi,\Env = \true$, where $\varphi$ is an \LTLf formula.
\end{theorem}

\begin{comment}
\begin{proof}
By Lemma \ref{lem:autconstruction}, the language of the \DA in Line 1 is $\L(\forall \varphi)$.
By Lemma~\ref{lem:games-synthesis} it is enough to find check $\iota$ is a winning state, and in this case to compute a winning strategy $f_\ag$ and return $\textsc{Strategy}(f_\ag)$. But finding the winning region and a winning strategy (for every state in the winning region) for safety games is done in Line 2 using Lemma~\ref{lem:solvereach}.
\end{proof}
\end{comment}

\section{Reachability and Safety Tasks, No Env Spec}
Algorithm \ref{alg:reachsafe} handles the case that $\Task$ is of the form $\exists \varphi_1 \land \forall \varphi_2$ where $\varphi_1$ and $\varphi_2$ are \LTLf formulas, and $\Env = \true$.

Intuitively, the algorithm proceeds as follows. First, it computes the corresponding \DA for $\forall \varphi_2$ and solves the safety game over it. The resulting winning area represents the set of states from which the agent has a strategy to realize its safety task. Then, it restricts the game area to the agent's winning area. Finally, it solves the reachability game over the game product of the corresponding \DA of $\exists \varphi_1$ and the remaining part of the \DA for $\forall \varphi_2$.

\begin{algorithm}[h!]
\caption{$\Task = \exists \varphi_1 \land \forall \varphi_2, \Env = true$ }
\label{alg:reachsafe}
\textbf{Input}: \LTLf formulas $\varphi_1$ and  $\varphi_2$ \\
\textbf{Output}: agent strategy $\sigma_\ag$ that realizes $\exists \varphi_1$ and $\forall \varphi_2$

\begin{algorithmic}[1]
%\label{alg:reachsafe}
\STATE $\A_1 = \textsc{ConvertDA}(\exists \varphi_1)$, say $\A_1 = (\D_{\exists \varphi_1},\reach(T_1))$ 
\STATE $\A_2 =  \textsc{ConvertDA}(\forall \varphi_2)$, say $\A_2 = (\D_{\forall \varphi_2},\safe(T_2))$

\STATE $(S_2,f_\ag)= \textsc{Solve}_{ag}(\A_2)$

% \STATE $(S_2,f_\ag)= $  \textsc{Solve$_{ag}$}($\D_{\forall \varphi_2},\safe(T_2))$

\STATE $\D_{\forall \varphi_2}'=$ \textsc{Restrict($\D_{\forall \varphi_2},S_2$)}, say the sink state is $\bot_2$
%STATE $\B =(\Sigma,Q_1 \times \Win_{ag},\delta_3,(q_1^0,q_2^0),Reach(T_1 \times \Win_{ag})$
\STATE  $\D =$ \textsc{Product($\D_{\exists \varphi_1},\D_{\forall \varphi_2}'$)}
\STATE $(R,g_\ag)= $  \textsc{Solve$_{ag}$}($\D, \reach(T_1 \times S_2))$
\STATE \textbf{if} {$\iota \not\in R$} \textbf{return} ``Unrealisable" \textbf{endif}
%\STATE $\sigma_\ag = \textsc{Strategy}(h_\ag=(g_\ag,f_\ag''))$
\STATE $h_\ag=$ $\textsc{Combine}(\D, R,g_\ag, f_\ag)$
%\COMMENT{$f_\ag''$ is the restriction of $f_\ag$ on  $\D_{\forall \varphi_2}$ }
\RETURN $\sigma_\ag=\textsc{Strategy}(\D,h_\ag)$

\end{algorithmic}
\end{algorithm}

\vspace{-0.3cm}

In order to obtain the final strategy for the agent we need to refine the strategy $\f_\ag$ to deal with the sink state, call it $\bot_2$, and combine it with $g_\ag$. Given $f_\ag$ computed in Line 3, define $f''_\ag:Q_1 \times (S_2 \cup \{\bot_2\}) \rightarrow 2^\Y$ over $\D$ by $f_\ag''(q,s) = f_\ag(s)$ if $s \in S_2$, and $f_\ag''(q,s) = Y$ (for some arbitrary $Y$) otherwise. In words, $f_\ag''$ ensures the second component stays in $S_2$ (and thus in $T_2$). Recall that $g_\ag$ over $\D$ ensures that $T_1$ is reached in the first co-ordinate, while at the same time maintaining the second co-ordinate is in $S_2$. Finally, let $\textsc{Combine}(\D, R, g_\ag,f_\ag)$ denote the final strategy $h_\ag: Q_1 \times (S_2 \cup \{\bot_2\}) \rightarrow 2^\Y$ defined as follows:  $h_\ag((q,s))=g_\ag((q,s))$ if $(q,s) \in R$, and   $h_\ag((q,s))=f_\ag''((q,s))$ otherwise. Intuitively, the agent following $h_\ag$ will achieve the reachability goal while staying safe, whenever this is possible, and stays safe otherwise.

\begin{theorem}\label{thm:reachsafe}
Algorithm~\ref{alg:reachsafe} solves synthesis under environment specifications problem with $\Task = \exists \varphi_1 \land \forall \varphi_2,\Env = \true$, where the $\varphi_i$ are \LTLf formulas.
\end{theorem}

% $f_\ag':S_2 \cup \{\bot\} \rightarrow 2^\Y$ over $\D_{\forall \varphi_2}'$ by $f_\ag'(q)=f_\ag(q)$ for $q \in S_2$, and $f_\ag'(\bot) = Y$ for an arbitrary $Y \in 2^\Y$; then, define $f_\ag'':Q_1 \times (S_2 \cup \{\bot\}) \rightarrow 2^\Y$ over $\D$ as $f_\ag''(q,s)=f_\ag'(s)$, for $(q,s) \in Q_1 \times (S_2 \cup \{\bot\})$. 
% %
% Therefore, the final strategy can be defined as the combination of $f_\ag''$ and $\g_\ag$.  Let 

\begin{comment}
\begin{proof}
By Lemma \ref{lem:autconstruction}, the language of the \DA in Line 1 is $\L(\exists \varphi_1)$, and the language of the \DA in Line 2 is $\L(\forall \varphi_2)$.
Let  $S_1$  be the agent winning region in the safety game over $\D_{\forall \varphi_2}$, we have that $S_1$ represents  the set of winning strategies for the agent enforcing $\forall \varphi_2$.
Restricting all the environment strategies to consider only those
that enforce $L(\forall \varphi_2)$ is done in Line 5 using Lemma \ref{lem:lift strategies to product}. By Lemma 2 it is enough to check $\iota$ is a winning state, and in this case to compute a winning strategy $f_\ag$ and return  $\textsc{Strategy}(f_\ag)$.
\end{proof}
\end{comment}

\section{Reachability Tasks, Safety Env Specs}

Algorithm \ref{alg:safe->reach} handles the case that $\Task$ is of the form $\exists \varphi_1$ and $\Env = \forall \varphi_2$, where $\varphi_1, \varphi_2$ are \LTLf formulas. A similar problem of this case was solved in \cite{DDTVZ21}, which, more specifically, considers only finite safety of the agent, i.e., the agent is required to stay safe until some point (the bound is related to an additional agent reachability task), and thus can actually be considered as reachability. 

Intuitively, the algorithm first computes all the environment strategies that can enforce $\Env = \forall \varphi_2$ \cite{BJW02}, represented as a restriction of the \DA for $\forall \varphi_2$, as in the previous section. Then, based on restricting the game arena on these environment strategies, the algorithm solves the reachability game over the product of the corresponding \DA of $\exists \varphi_1$ and the restricted part of the \DA for $\forall \varphi_2$. 

\begin{algorithm}[h]
\caption{$\Task = \exists \varphi_1, \Env = \forall \varphi_2$}
\label{alg:safe->reach}
\begin{algorithmic}[1]
\REQUIRE \LTLf formulas $\varphi_1, \varphi_2$
\ENSURE agent strategy $\sigma_\ag$ that enforces $\exists \varphi_1$ under $\forall \varphi_2$
% \STATE 
% $(\D_1, \reach(T_1)) = \textsc{ConvertDA}(\exists \varphi_1)$
% \STATE 
% $(\D_2, \safe(T_2)) = \textsc{ConvertDA}(\forall \varphi_2)$

%\changed{
\STATE $\A_1 = \textsc{ConvertDA}(\exists \varphi_1)$, say $\A_1 = (\D_{\exists \varphi_1},\reach(T_1))$ 
\STATE $\A_2 =  \textsc{ConvertDA}(\forall \varphi_2)$, say $\A_2 = (\D_{\forall \varphi_2},\safe(T_2))$
%}
\STATE 
$(S_2, f_\env)=   \textsc{Solve}_\env(\A_2)$ %($\D_2, \safe(T_2)$)
\STATE 
$\D'_2 = \textsc{Restrict}(\D_2, S_2)$, say the sink state is $\bot_2$
\STATE 
$\D = \textsc{Product}(\D_1, \D'_2)$
\STATE 
$(R, f_\ag)= \textsc{Solve}_{\ag}(\D, \reach((T_1 \times S_2) \cup (Q_1 \times \{\bot_2\}))$
\STATE \textbf{if} {$\iota \not\in R$} \textbf{return} ``Unrealisable" \textbf{endif}

%\IF {$\iota \not\in R_\ag$}
%\RETURN ``Unrealisable"
%\ENDIF
%\STATE $\sigma_\ag = \textsc{Strategy}(f_\ag)$
\RETURN $\sigma_\ag = \textsc{Strategy}(\D,\f_\ag)$
\end{algorithmic}
\end{algorithm}

\vspace{-0.2cm}
\begin{theorem}\label{thm:alg-safe->reach}
Algorithm~\ref{alg:safe->reach} solves the synthesis under environment specifications problem with $\Task = \exists \varphi_1, \Env = \forall \varphi_2$, where the $\varphi_i$ are \LTLf formulas.
\end{theorem}
\begin{comment}
\begin{proof}
By Lemma~\ref{lem:autconstruction}, the language of the \DA $A_1 = (\D_1, \reach,T_1)$ in Line 1 is $\L(\exists \varphi_1)$ and the language of the \DA $A_2 = (\D_2, \safe,T_2)$ in Line 2 is $\L(\forall \varphi_2)$. Finding all the environment strategies that enforce $\A_2$ through safety games considering the environment as the protagonist is done in Lines 3\&4 using Lemma~\ref{lem:solvesafe} and Remark~\ref{rm:allsafestrategies}. Restricting all the environment strategies to considering only those that enforce $\A_2$ is done in Line 5 using Lemma~\ref{lem:lift strategies to product}. By Lemma~\ref{lem:games-synthesis} it is enough to find check $\iota$ is a winning state, and in this case to compute a winning strategy $f_\ag$ and return $\textsc{Strategy}(f_\ag)$. But finding the winning region and a winning strategy (for every state in the winning region) for reachability games is done in Line 6 using Lemma~\ref{lem:solvereach}.
\end{proof}
\end{comment}

\section{Safety Tasks, Safety Env Specs}
Algorithm \ref{alg:safe->safe} handles the case that $\Task$ is of the form $\forall \varphi_1$ and $\Env = \forall \varphi_2$, where $\varphi_1, \varphi_2$ are \LTLf formulas. 

Intuitively, the algorithm proceeds as follows. First, it computes the corresponding \DA for $\forall \varphi_2$ and solves the safety game for the environment over it. The resulting winning area represents the set of states, from which the environment has a strategy to enforce the environment specification $\L(\forall \varphi_2)$. It is worth noting that restricting the \DA to considering only such winning area, in fact, captures all the environment strategies that enforce $\L(\forall \varphi_2)$~\cite{BernetJW02}. Based on the restriction, the algorithm solves the safety game over the product of the corresponding \DA of $\forall \varphi_1$ and the remaining part of the \DA for $\forall \varphi_2$.

\begin{algorithm}[ht]
\caption{$\Task = \forall \varphi_1, \Env = \forall \varphi_2$}
\label{alg:safe->safe}
\begin{algorithmic}[1]
\REQUIRE \LTLf formulas $\varphi_1, \varphi_2$
\ENSURE agent strategy $\sigma_\ag$ that enforces $\forall \varphi_1$ under $\forall \varphi_2$
\STATE 
$\A_1 = \textsc{ConvertDA}(\forall \varphi_1)$, say $\A_1 = (\D_1, \safe(T_1))$
\STATE 
$\A_2= \textsc{ConvertDA}(\forall \varphi_2)$, say $\A_2 = (\D_2, \safe(T_2))$
\STATE 
$(S_2, f_\env)= \textsc{Solve}_\env(\A_2)$
\STATE 
$\D'_2 = \textsc{Restrict}(\D_2, S_2)$, call the sink $\bot_2$
\STATE 
$\D = \textsc{Product}(\D_1, \D'_2)$
\STATE 
$(S, f_\ag)= \textsc{Solve}_{\ag}(\D, \safe((T_1 \times S_2) \cup (Q_1 \times \{\bot_2\})))$
\STATE \textbf{if} {$\iota \not\in S$} \textbf{return} ``Unrealisable" \textbf{endif}
%\IF {$\iota \not\in S_\ag$}
%\RETURN ``Unrealisable"
%\ENDIF
%\STATE $\sigma_\ag = \textsc{Strategy}(f_\ag)$
\RETURN $\sigma_\ag = \textsc{Strategy}(\D,f_\ag)$
\end{algorithmic}
\end{algorithm}
%\vspace{-0.3cm}

\begin{theorem}\label{thm:alg-safe->safe}
Algorithm~\ref{alg:safe->safe} solves the synthesis under environment specifications problem with $\Task = \forall \varphi_1, \Env = \forall \varphi_2$, where the $\varphi_i$ are \LTLf formulas.
\end{theorem}
\begin{comment}
\begin{proof}
By Lemma~\ref{lem:autconstruction}, the language of the \DA $A_1 = (\D_1, \safe, T_1)$ in Line 1 is $\L(\forall \varphi_1)$ and the language of the \DA $A_2 = (\D_2, \safe, T_2)$ in Line 2 is $\L(\forall \varphi_2)$. Finding all the environment strategies that enforce $\L(\A_2)$ through safety games considering the environment as the protagonist is done in Lines 3\&4 using Lemma~\ref{lem:solvesafe} and Remark~\ref{rm:allsafestrategies}. Restricting all the environment strategies to considering only those that enforce $\L(\A_2)$ is done in Line 5 using Lemma~\ref{lem:lift strategies to product}. By Lemma~\ref{lem:games-synthesis} it is enough to find check $\iota$ is a winning state, and in this case to compute a winning strategy $f_\ag$ and return $\textsc{Strategy}(f_\ag)$. But finding the winning region and a winning strategy (for every state in the winning region) for safety games is done in Line 6 using Lemma~\ref{lem:solvesafe}.
\end{proof}
\end{comment}

\section{Reachability and Safety Tasks, Safety Env Specs}

Algorithm~\ref{alg:safe->reach+safe} handles the case that $\Task$ is of the form $\exists \varphi_1 \land \forall \varphi_2$ and $\Env = \forall \varphi_3$, where $\varphi_1, \varphi_2, \varphi_3$ are \LTLf formulas. As mentioned in the previous section, a similar problem of this case that considers only finite safety of the agent was solved in \cite{DDTVZ21} by reducing $\Task$ to reachability properties only. Instead, we provide here an approach to the synthesis problem considering infinite agent safety.

Intuitively, the algorithm proceeds as follows. Following the algorithms presented in the previous sections, it first computes all the environment strategies that can enforce $\Env = \varphi_3$, represented as a restriction of the \DA for $\forall \varphi_3$. Then, based on restricting the game arena on these environment strategies, the algorithm solves the safety game for the agent over the product of the corresponding \DA of $\forall \varphi_2$ and the restricted part of the \DA for $\forall \varphi_3$. This step is able to capture all the agent strategies that can realize $\forall \varphi_2$ under environment specification $\forall \varphi_3$. Next, we represent all these agent strategies by restricting the product automaton to considering only the computed agent winning states, thus obtaining $\D'$. Finally, the algorithm solves the reachability game over the product of the corresponding \DA of $\exists \varphi_1$ and $\D'$. In order to abstract the final strategy for the agent, it is necessary to combine the two agent strategies: one is from the safety game for enforcing $\forall \varphi_2$ under $\forall \varphi_3$, the other one is from the final reachability game for enforcing $\exists \varphi_1$ while not violating $\forall \varphi_2$ under $\forall \varphi_3$.

\begin{algorithm}
\caption{$\Task =\exists \varphi_1 \land \forall \varphi_2, \Env = \forall \varphi_3$}
\label{alg:safe->reach+safe}
\begin{algorithmic}[1]
\REQUIRE \LTLf formulas $\varphi_1, \varphi_2, \varphi_3$
\ENSURE agent strategy $\sigma_\ag$ that enforces $\exists 
\varphi_1 \land \forall \varphi_2$ under $\forall \varphi_3$

\STATE 
%changed{
$\A_1 = \textsc{ConvertDA}(\exists \varphi_1)$, say $\A_1 = (\D_1, \reach(T_1))$
%}
\STATE 
%\changed{
$\A_2 =  \textsc{ConvertDA}(\forall \varphi_2)$, say $\A_2 = (\D_2, \safe(T_2))$
%}
\STATE 
%\changed{
$\A_3 =  \textsc{ConvertDA}(\forall \varphi_3)$, say $\A_3 = (\D_3, \safe(T_3))$
%}
\STATE 
%\changed{
$(S_3, f_\env)=  \textsc{Solve}_\env(\A_3)$ %$\D_3, \safe(T_3))$
%}
\STATE 
%\changed{
$\D'_3 = \textsc{Restrict}(\D_3, S_3)$, call the sink $\bot_3$
%}
\STATE 
$\D = \textsc{Product}(\D_2, \D'_3)$
\STATE 
%\changed{
$(S_2, f^s_\ag)= $   \textsc{Solve$_\ag$}($\D, \safe((T_2 \times S_3) \cup (Q_2 \times \{\bot_3\}))$)
%}
% \IF {$q^{\D}_0 \not\in S$}
% \RETURN ``Unrealisable"
% \ENDIF
\STATE 
%\changed{
$\D' = \textsc{Restrict}(\D, S_2)$, call the sink $\bot_2$
%}
\STATE $\C = \textsc{Product}(\D_1, \D')$
\STATE Let $f^{s'}_\ag: Q_1 \times (S_2 \cup \{\bot_2\}) \rightarrow 2^\Y$ map $(q_1,q_2)$ to $f^s_\ag(q_2)$ if $q_2 \in S_2$, and is arbitrary otherwise.
\COMMENT{$f^{s'}_\ag$ lifts $f^s_\ag$ to  $\C$}
\STATE 
% $(R, f^r_\ag)= \textsc{Solve}_{\ag}(\C, \reach((T_1 \times S_2 \times S_3) \cup (Q_1 \times (S_2 \cup \{\bot_2\})\times \{\bot_3\} ))$ 
{$(R, f^r_\ag)= \textsc{Solve}_{\ag}(\C, \reach((T_1 \times S_2 \times S_3) \cup (Q_1 \times  (\eta(S_2) \cup \{\bot_2\}) \times \{\bot_3\} ))$ }\\
\COMMENT{ $\eta:Q_2 \times Q_3 \to Q_2$ is the projection onto $Q_2$, i.e., $(q_2,q_3) \mapsto q_2$}
\STATE \textbf{if} {$\iota \not\in R$} \textbf{return} ``Unrealisable" \textbf{endif}
%\IF {$q^{\C}_0 \not\in R$}
%\RETURN ``Unrealisable"
%\ENDIF
\STATE Let $f_\ag:Q_1 \times (S_2 \cup \{\bot_2\}) \rightarrow 2^\Y$ on $\C$ map $q$ to $f^r_\ag(q)$ if $q \in R$, and to $f^{s'}_\ag(q)$ otherwise.
\COMMENT{$f_\ag$ does $f^r_\ag$ on $R$, and $f^{s'}_\ag$ otherwise.}
% \STATE $f_\ag = \textsc{Combine}(\C, R, f^{s'}_\ag, f^r_\ag)$  \\
\RETURN $\sigma_\ag = \textsc{Strategy}(\C, f_\ag)$
\end{algorithmic}
\end{algorithm}
%\vspace{-0.3cm}

\begin{theorem}\label{thm:alg-safe->reach+safe}
Algorithm~\ref{alg:safe->reach+safe} solves synthesis under environment specifications problem with $\Task = \exists \varphi_1 \land \forall \varphi_2, \Env = \forall \varphi_3$, where the $\varphi_i$ are \LTLf formulas.
\end{theorem}

\section{Reachability and Safety Tasks and Env Specs}

Algorithm~\ref{alg:reach+safe->reach+safe} handles the case that $\Env = \forall \varphi_1 \land \exists \varphi_2$ and $\Task = \exists \varphi_3 \land \forall \varphi_4$ by solving synthesis for the formula $\Env \limp \Task$ \cite{AGMR19}, i.e., for
$
(\exists \lnot \varphi_1 \lor \forall \lnot \varphi_2) \lor (\exists \varphi_3 \land \forall \varphi_4).
$
Note that, from the general case, we get all cases involving reachability environment specifications by suitably setting $\varphi_1, \varphi_2$ or $\varphi_4$ to $true$. We remark that for the case $\varphi_4 = \true$ in which the safety and reachability specifications are presented in the safety-fragment and co-safety fragment of \LTL is solved in~\cite{CamachoBM19}.

We first define two constructions that will be used in the algorithm.  Given a transition system $\D = (\Sigma, Q, \iota, \delta)$ and a set of states $T \subseteq Q$, define  $\textsc{Flagged}(\D, T)$ to be the transition system that, intuitively, records whether a state in $T$ has been seen so far. Formally, $\textsc{Flagged}(\D, T)$ returns the transition system $D^f = (\Sigma, Q^f, \iota^f, \delta^f)$ defined as follows:
\begin{enumerate*}
\item 
$Q^f = Q \times \{yes,no\}$.
\item 
$\iota^f = (\iota,b)$, where $b = no$ if $\iota \not \in T$, and $b = yes$ if $\iota \in T$.
\item
$\delta^f((q,b),z) = (q',b')$ if $\delta(q,z)=q'$ and one of the following conditions holds: (i) $b = b' = yes$, (ii) $b = b' = no, q' \not \in T$, (iii) $b = no, b' = yes, q' \in T$.
\end{enumerate*}
Given a transition system $\D = (\Sigma, Q, \iota, \delta)$ and disjoint subsets $V_0,V_1$ of $Q$, define $\textsc{RestrictionWithSinks}(\D,V_0,V_1)$ to be the transition system on state set $V_0$ that, intuitively, behaves like $\D$ on $V_0$, transitions from $V_0$ to $V_1$ are redirected to a new sink state $\bot$, and transitions from $V_0$ to $Q \setminus (V_0 \cup V_1)$ are redirected to a new sink state $\top$. Formally, $\textsc{RestrictionWithSinks}(\D,V_0,V_1)$ is the transition system $(\Sigma, \hat{Q}, \hat{\iota}, \hat{\delta})$ defined as follows:
\begin{enumerate*}
    \item 
    $\hat{Q} = V_0 \cup \{\top,\bot\}$.
    \item 
    $\hat{\iota} = \iota$.
    \item
    $\hat{\delta}(q, z) = \delta(q, z)$ if $\delta(q, z) \in V_0$. Otherwise, define $\hat{\delta}(q,z) = \bot$ if $\delta(q, z) \in V_1$, and $\hat{\delta}(q,z) = \top$ if $\delta(q, z) \in Q \setminus (V_0 \cup V_1)$.
\end{enumerate*}

\begin{algorithm}[!h]
\caption{$\Task =\exists \varphi_3 \land \forall \varphi_4, \Env = \forall \varphi_1 \land \exists \varphi_2$}
\label{alg:reach+safe->reach+safe}
\begin{algorithmic}[1]
\REQUIRE \LTLf formulas $\varphi_1, \varphi_2, \varphi_3, \varphi_4$
\ENSURE agent strategy $\sigma_\ag$ that enforces $\exists \varphi_3 \land \forall \varphi_4$ under $\forall \varphi_1 \land \exists \varphi_2$
\STATE 
{
$\A_1 =  \textsc{ConvertDA}(\exists \lnot \varphi_1)$, say $\A_1 = (\D_1, \reach(B_1))$
\STATE 
$\A_2 = \textsc{ConvertDA}(\forall \lnot \varphi_2)$, say $\A_2 = (\D_2, \safe(B_2))$
\STATE 
$\A_3 = \textsc{ConvertDA}(\exists \varphi_3)$, say $\A_3 = (\D_3, \reach(B_3))$
\STATE 
$\A_4 = \textsc{ConvertDA}(\forall \varphi_4)$, say $\A_4 = (\D_4, \safe(B_4))$
}
\STATE 
$\D_p = \textsc{Product}(\D_1, \D_2, \D_3, \D_4)$ 
\STATE Let $Q_p$ be the state set of $\D_p$, and $T_i$ the lift of $B_i$ to $Q_p$ (for $i \leq 4$)
\STATE 
$(R_1, f^1_\ag)= \textsc{Solve}_\ag(\D_p, \reach(T_1))$ 
\STATE
$\D_p' = \textsc{Restrict}(\D_p, Q\setminus R_1)$
\STATE 
$\D^f = \textsc{Flagged}(\D_p', T_3)$
\STATE 
$(S_2, f^2_\ag)= \textsc{Solve}_\ag(\D^f, \safe( T_2))$ 
\STATE 
$(S_4, f^4_\ag)= \textsc{Solve}_\ag(\D^f, \safe(T_4))$ 
% \STATE 
% $\D' = \textsc{Restrict}(\D, S_4)$
\STATE 
$(R_3, f^3_\ag)= \textsc{Solve}_{\ag}(\textsc{Restrict}(\D^f, S_4), \reach(T_3))$ 
\STATE 
$V_0 = (Q^f \setminus (S_2 \cup S_4)) \cup ((S_4 \cap T_2) \setminus (R_3 \cup S_2))$ 
\STATE
$V_1$ is all states in $(S_4 \setminus T_2) \setminus (R_3 \cup S_2)$ whose flag is set to $no$
\STATE 
$\hat{\D} = \textsc{RestrictionWithSinks}(\D^f, V_0, V_1)$ 
\STATE 
$(E,f^e_\ag) = \textsc{Solve}_\ag(\hat{\D},\safe((T_2 \cap T_4) \cup \{\top\}))$

\STATE 
$W_\ag = S_2 \cup R_3 \cup E$ \COMMENT{Note that $W_\ag \subseteq Q^f \cup \{\top\}$}
\STATE \textbf{if} {$\iota \not\in W_\ag$} \textbf{return} ``Unrealisable" \textbf{endif}
%\IF {$\iota \not\in W_\ag$}
%\RETURN ``Unrealisable"
%\ENDIF
%
{\STATE $f_\ag = \textsc{Combine}(\D^f,$ $ f^1_\ag, f^2_\ag, f^3_\ag, f^4_\ag, R_1, S_2,R_3,E)$ \COMMENT{See the definition below.}}
\RETURN $\sigma_\ag = \textsc{Strategy}(\D^f, f_\ag)$

\end{algorithmic}
%\vspace{-0,1cm}

\end{algorithm}
%\vspace{-0.3cm}

Intuitively, at Line 10, $S_2$ will form part of the agent's winning region since from here $\safe(T_2)$ can be ensured. At Line 12, $R_3$ will also form part of the agent's winning region since from $R_3$ in $\D'$ $\reach(T_3) \cap \safe(T_4)$ can be ensured. In the following steps, we identify remaining ways that the agent can win, intuitively by maintaining $T_2 \cap T_4$ either forever (in which case $\safe(T_2)$ is ensured), or before the state leaves $T_2 \cap T_4$ either (i) it is in $S_2$ or $R_3$ (in which case we proceed as before), or otherwise (ii) it is in $S_4$ (but not in $S_2$ nor in $R_3$) and has already seen $T_3$ (in which case $\reach(T_3) \cap \safe(T_4)$ can be ensured).

At the end of the algorithm, we combine the four strategies $f^1_\ag, f^2_\ag, f^3_\ag$ and $f^4_\ag$ through procedure $\textsc{Combine}(\D^f,$ $ f^1_\ag, f^2_\ag, f^3_\ag, f^4_\ag, R_1,S_2,R_3,E)$ to obtain the final strategy $f_\ag: (Q^f)^+ \rightarrow 2^\Y$ as follows. For every history $h \in (Q^f)^+$, if the history ever enters $R_1$ then follow $f^1_\ag$, ensuring $\reach(T_1)$, otherwise, writing $q$ for the start state of $h$:
% Basically, V_0 is everything without S_2 or R_3, and with a little part of S_4 (which without V_1 is just S_4 \setminus R_3,T_2 with flag "yes"
\begin{enumerate*}
    \item 
    if $q \in S_2$ then use $f^2_\ag$, which ensures $\safe(T_2)$;
    \item 
    if $q \in R_3$ then use $f^3_\ag$ until $T_3$ is reached and thereafter use $f^4_\ag$, which ensures $\safe(T_4) \cap \reach(T_3)$;
    \item 
    if $q \in E$ then use $f^e_\ag$ while the states are in $E$, ensuring $\safe(T_2)$ if play stays in $E$; if ever, let $q'$ be the first state in the history that is not in $E$; by construction, this corresponds to $\top$ in $\D^f$ and thus is (i) in $S_2$ or (ii) in $R_3$, and so proceed as before, or else (iii) in $(S_4 \setminus T_2) \setminus (R_3 \cup S_2)$ (which can be simplified to $S_4 \setminus (R_3 \cup T_2)$) with flag value $yes$ in which case  switch to strategy $f^4_\ag$. Intuitively, case (i) ensures $\safe(T_2)$, and cases (ii) and (iii) each ensure $\safe(T_4) \cap \reach(T_3)$;
    \item and if none of these, then make an arbitrary move.
\end{enumerate*}
{Note that in spite of being a function of the whole history, $f_\ag$ can be represented by a finite-state transducer. So in the Algorithm 7, as before, with a little abuse of notation we write directly $\sigma_\ag = \textsc{Strategy}(\D^f, f_\ag)$, to mean that we return its representation as a transducer.}

\begin{theorem}~\label{thm:reach+safe->reach+safe}
Algorithm~\ref{alg:reach+safe->reach+safe} solves the synthesis under environment specifications problem with $\Task = \exists\varphi_3 \land \forall \varphi_4$ and $\Env = \forall \varphi_1 \land \exists \varphi_2$, where  $\varphi_i$ are \LTLf formulas.
\end{theorem}

\paragraph{Comparison to Algorithms 1-6.} Note that Algorithm~\ref{alg:reach+safe->reach+safe} can solve the other six variants by suitably instantiating some of $\varphi_1, \varphi_2, \varphi_3, \varphi_4$ to $true$. Nevertheless, Algorithm~\ref{alg:reach+safe->reach+safe} is much more sophisticated than Algorithms 1-6. Hence, in this paper, we present the algorithms deductively, starting with simpler variants and moving to the most difficult. Furthermore, instantiating Algorithm~\ref{alg:reach+safe->reach+safe} does not always give the same algorithms as Algorithms 1-6. For instance, Algorithm~\ref{alg:reach} for the synthesis problem of $\Task = \exists \varphi$~(no environment specification) can be obtained from Algorithm~\ref{alg:reach+safe->reach+safe} by setting $\varphi_1, \varphi_2, \varphi_4$ to $true$, but we cannot get Algorithm~\ref{alg:safe->reach} for the synthesis problem of $\Env = \forall \varphi$ and $\Task = \exists \psi$ in this way. This is because Algorithm~\ref{alg:reach+safe->reach+safe} solves the synthesis problem by reducing to $\Env \rightarrow \Task$ \cite{AGMR19}, but Algorithm~\ref{alg:safe->reach} directly disregards all environment strategies that cannot enforce $\Env$ by first solving a safety game for the environment on $\Env$ and removing all the states that do not belong to the environment winning region to get a smaller game arena, hence obtaining optimal complexity. Analogously, in Algorithm~\ref{alg:reachsafe} for the synthesis problem of $\Env = true$ and $\Task = \exists \varphi_1 \land \forall \varphi_2$, we also first disregard all the agent strategies that are not able to enforce $\forall \varphi_2$, obtaining a smaller game arena for subsequent computations, hence getting an optimal complexity in practice compared to constructing the game arena considering the complete state space from the DA of $\forall \varphi_2$.

\section{Conclusion}
In this paper, we have studied the use of reachability and safety properties based on \LTLf for both agent tasks and environment specifications.
As mentioned in the introduction, though we have specifically focused  on \LTLf, all algorithms presented here can be readily applied to other temporal logics on finite traces, such as Linear Dynamic Logics on finite traces~(\LDLf), which is more expressive than \LTLf \cite{DegVa13}, and Pure-Past \LTL~\cite{DDFR20}, as long as there exists a technique to associate formulas to equivalent \DFAs. % Extensions to other temporal logics, such as the mu-calculus with past modalities, 

It is worth noting that all the cases studied here are specific Boolean combinations of $\exists\varphi$. It is of interest to indeed devise algorithms to handle arbitrary Boolean combinations. Indeed, considering that \LTLf is expressively equivalent to pure-past \LTL, an arbitrary Boolean combination of $\exists\varphi$ would correspond to a precise class of \LTL properties in Manna \& Pnueli's Temporal Hierarchy \cite{MannaPnueli90}: the so-called \emph{obligation} properties. We leave this interesting research direction for future work.

Another direction is to consider best-effort synthesis under assumptions for Boolean combinations of $\exists \varphi$, instead of (ordinary) synthesis under assumptions, in order to handle ignorance the agent has about the environment~\cite{DBLP:conf/ijcai/AminofGR21,DBLP:conf/kr/AminofGLMR21,DBLP:conf/lics/AminofGRZ23,DBLP:conf/stacs/Berwanger07,DBLP:conf/mfcs/Faella09}.

\bibliographystyle{plain}
\bibliography{references,ref}

\end{document}